\documentclass[twocolumn]{aastex63}
\usepackage[utf8]{inputenc}
\usepackage{epsfig}
\usepackage{amsmath}

\shortauthors{J.-T. Li et al.}
\shorttitle{NGC~3079 superbubble}

\begin{document}

\title{Pressure Balance and Energy Budget of the Nuclear Superbubble of NGC~3079}

\correspondingauthor{Jiang-Tao Li}
%\email{jiangtaoli@pmo.ac.cn}
\email{pandataotao@gmail.com}
\correspondingauthor{Wei Sun}
\email{sunwei@pmo.ac.cn}

\author[0000-0001-6239-3821]{Jiang-Tao Li}
\affiliation{Purple Mountain Observatory, Chinese Academy of Sciences, 10 Yuanhua Road, Nanjing 210023, People’s Republic of China}

\author[0000-0002-5456-0447]{Wei Sun}
\affiliation{Purple Mountain Observatory, Chinese Academy of Sciences, 10 Yuanhua Road, Nanjing 210023, People’s Republic of China}

\author[0000-0001-7500-0660]{Li Ji}
\affiliation{Purple Mountain Observatory, Chinese Academy of Sciences, 10 Yuanhua Road, Nanjing 210023, People’s Republic of China}

\author[0000-0001-7254-219X]{Yang Yang}
\affiliation{Purple Mountain Observatory, Chinese Academy of Sciences, 10 Yuanhua Road, Nanjing 210023, People’s Republic of China}

\begin{abstract}
Superbubbles in the nuclear region of galaxies could be produced by the AGN or nuclear starburst via different driving forces. We report analysis of the multi-wavelength data of the kpc-scale nuclear superbubble in NGC~3079, in order to probe the mechanisms driving the expansion of the superbubble. Based on the \emph{Chandra} X-ray observations, we derive the hot gas thermal pressure inside the bubble, which is about one order of magnitude higher than that of the warm ionized gas traced by optical lines. We derive a [\ion{C}{2}]-based star formation rate of ${\rm SFR}\sim1.3\rm~M_\odot~{\rm yr}^{-1}$ from the nuclear region using the \emph{SOFIA}/FIFI-LS observation. This SFR infers a radiation pressure toward the bubble shells much lower than the thermal pressure of the gases. The \emph{VLA} radio image infers a magnetic pressure at the northeast cap above the superbubble less than the thermal pressure of the hot gas enclosed in the bubble, but has a clearly larger extension. The magnetic field may thus still help to reconcile the expansion of the bubble. The observed thermal energy of the hot gas enclosed in the bubble requires an energy injection rate of $\gtrsim10^{42}\rm~ergs~s^{-1}$ within the bubble's dynamical age, which is probably larger than the power provided by the current nuclear starburst and the parsec-scale jet. If this is true, stronger past AGN activity may provide an alternative energy source to drive the observed bubble expansion.
\end{abstract}


\section{Introduction} \label{sec:intro}

Superbubbles in the nuclear regions of galaxies could be produced either by nuclear starburst or AGN. A well known example of such a galactic nuclear superbubble is the ``Fermi bubble'' in our Milky Way (MW) galaxy \citep{Su2010} and its multi-wavelength counterparts (e.g., \citealt{BlandHawthorn03,BH2019,Finkbeiner04,Predehl20}), for which various models based on star formation or AGN feedback have been proposed (e.g., \citealt{Guo2012,BH2013,Crocker2015}). Understanding the physical processes involved in driving the superbubble via studies of the ``Fermi bubble'' is limited by the uncertainty of past activities in the currently quiescent Galactic nuclear region. Models can produce the bubble assuming a huge range of energy input rates and timescales (e.g., \citealt{Guo2012}). It is therefore complementary to study an analog of the ``Fermi bubble'' at an earlier stage, in an external galaxy with ongoing nuclear starburst and/or AGN.

A galactic-scale outflow can be driven by different mechanisms, such as the thermal pressure \citep[e.g.,][]{Chevalier1985}, radiation pressure \citep[e.g.,][]{Krumholz2012}, momentum (e.g., \citealt{Murray2005}), cosmic rays (CRs; e.g., \citealt{Heesen2018}), and/or MHD waves (e.g., \citealt{Breitschwerdt1991,Krumholz2012}). In most of the studies of external galaxies, we still have few observational constraints on what creates the galactic nuclear superbubbles and how they expand outward. For example, it is still unclear how the bubble distributes its energy into different phases of the interstellar medium (ISM), in the form of kinetic energy, thermal energy of gas at different temperatures and densities, radiation, CRs, and magnetic field, etc. The degree of pressure balance between them are key probes of how the expansion of the bubble is driven. We need high-quality multi-wavelength data to estimate the energy densities contained in these different ISM phases. 

We herein present a multi-wavelength study of NGC~3079 ($d=20.6\rm~Mpc$, $1^{\prime\prime}=100\rm~pc$), which hosts a $1.5\rm~kpc$-diameter nuclear superbubble bright in radio, H$\alpha$, and soft X-ray (e.g., Fig.~\ref{fig:MultiColorImage}; \citealt{Duric1988,Veilleux1994,Cecil2002}). Compared to the ``Fermi bubble'' in our Galaxy, a key advantage of this superbubble is that it is much smaller. Assuming a typical outflow velocity of $\sim500\rm~km~s^{-1}$, we obtain a dynamical timescale of $\sim$1.5~Myr for the bubble (see more discussions on the bubble age in the following sections), comparable to the lifetime of a $30\rm~M_\odot$ star. This means that everything we have witnessed in the bubble must be produced in situ, and directly related to the energy sources in the nuclear region (starburst or AGN; e.g., \citealt{Li2019}). Furthermore, an AGN and starburst hybrid scenario is also proposed to excite the dense gas tracers in NGC~3079 (e.g., \citealt{PB2007,LiFei2019MNRAS}).

The major goal of this paper is to examine the dominant driving force and energy budget of the superbubble in NGC~3079. In \S\ref{subsec:xray}, we present an analysis of the \emph{Chandra} observations (more details in Appendix~\ref{sec:ChandraSpecAnal}) and an estimate of the magnetic field strength from the radio data. We estimate the star formation rate (SFR) of the nuclear region based on the extinction free [\ion{C}{2}]~$\lambda$158$\mu$m line measured with the \emph{SOFIA} observation in \S\ref{subsec:fir}. We then discuss the pressure balance of different ISM phases in \S\ref{subsec:pres_comp}, and the energy budget of the superbubble in \S\ref{subsec:dyn}. Our main conclusions are summarized in \S\ref{sec:con}. Throughout the paper, all errors are quoted at 90\% confidence level.

\section{Data Reduction and Analysis} \label{sec:obs}

\subsection{Chandra and VLA}\label{subsec:xray}

\begin{figure*}[!th]
\begin{center}
\epsfig{figure=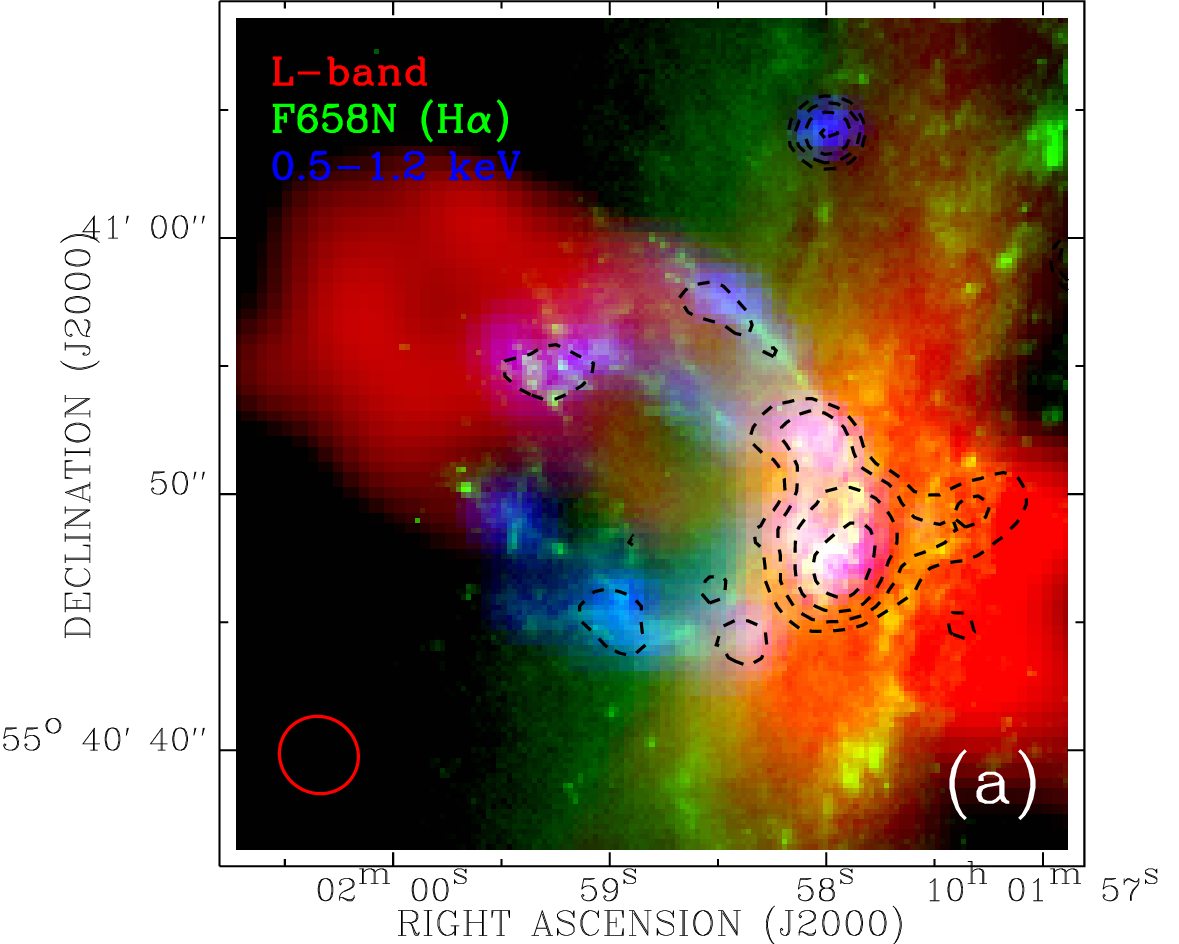,width=0.48\textwidth,angle=0, clip=}
\epsfig{figure=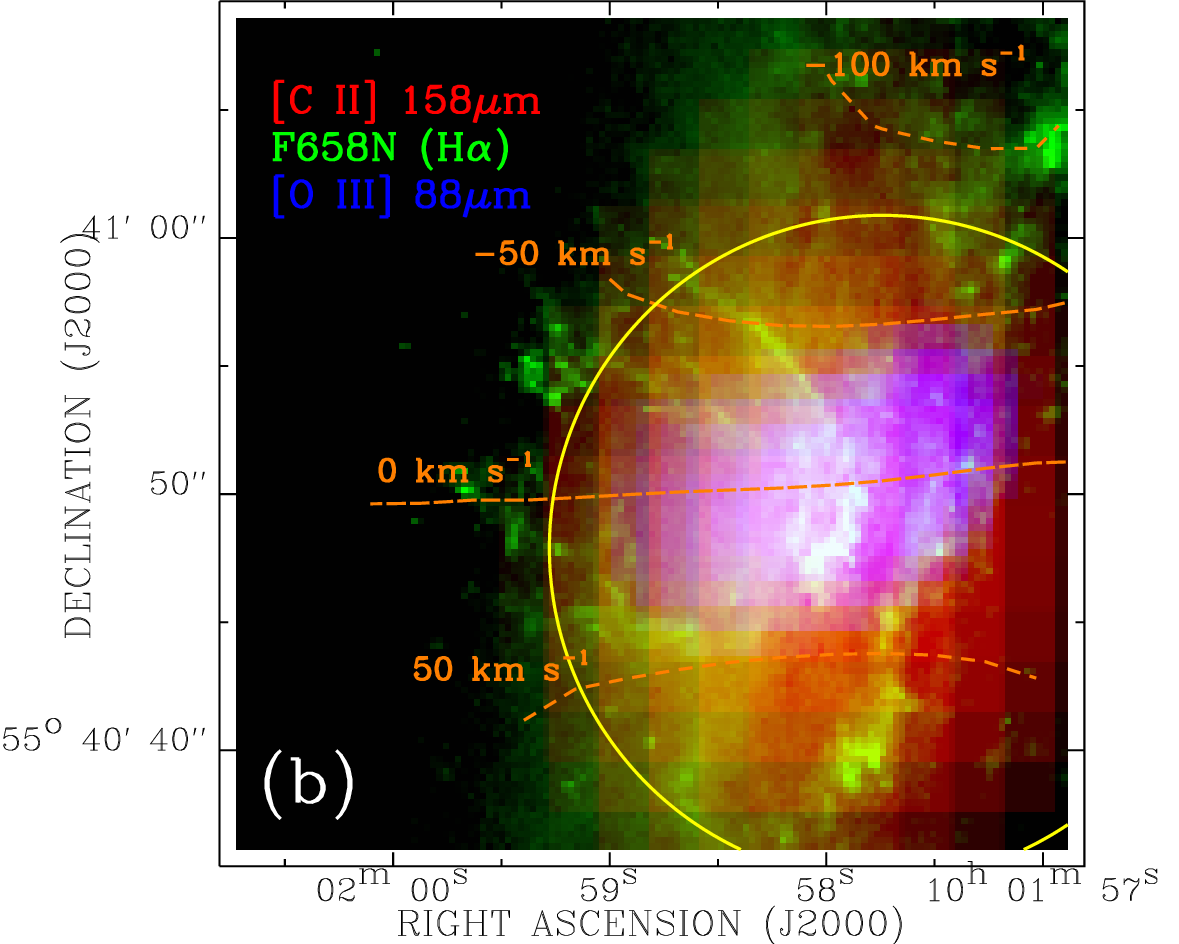,width=0.48\textwidth,angle=0, clip=}
\caption{Multi-band tri-color images of the NE nuclear superbubble of NGC~3079: (a) \emph{VLA} L-band (red), \emph{HST}/F658N (H$\alpha$, green), and \emph{Chandra} 0.5--1.2~keV (blue) images; (b) \emph{SOFIA} [\ion{C}{2}]~$\lambda157\mu$m line (red), \emph{HST}/F658N (green), and \emph{SOFIA} [\ion{O}{3}]~$\lambda88\mu$m line (blue) images. The images in those two panels cover the same field of view. In Panel~(a), the black contours outline the smoothed \emph{Chandra} 1.2--2.0~keV image at levels of (2, 4, 8, and 16)$\times10^{-4}$~photons~s$^{-1}$~cm$^{-2}$~arcmin$^{-2}$, and the red ellipse at the lower left corner shows the synthesized beam of the \emph{VLA} L-band observation. In Panel~(b), the orange dashed contours show the [\ion{C}{2}] line velocities with respect to the heliocentric velocity of NGC~3079  ($v_\odot=1116$~km~s$^{-1}$), and the yellow circle of $r=13^{\prime\prime}$ is used to calculate the SFR in \S\ref{subsec:fir}.} \label{fig:MultiColorImage}
\end{center}
\end{figure*}

In this paper, we use three of the four \emph{Chandra} archival observations covering NGC~3079 (ObsID 2038, 19307, 20947; ObsID 7851 is a snapshot observation with an effective exposure time of only 4.68~ks), with standard data calibration steps similar as those described in \citet{Li2019} (the \emph{VLA} radio image presented in Fig.~\ref{fig:MultiColorImage}a is also described in this reference). The total effective exposure time is 124.2~ks after filtering out the abnormal count rate period. The exposure corrected images are created using \emph{merge\_obs} with a bin size of 0.25 pixel (0.123\arcsec), adopting the Energy-Dependent Subpixel Event Repositioning (EDSER) algorithm \citep{Li2004}. The final background subtracted and adaptively smoothed images in the 0.5--1.2~keV and 1.2--2.0~keV bands are presented in Fig.~\ref{fig:MultiColorImage}a.

As shown in Fig.~\ref{fig:MultiColorImage}, the 0.5--1.2~keV emission from the superbubble shows some apparent coherent structures as the warm ionized gas traced by the H$\alpha$ emission \citep[see also in ][]{Cecil2002,Li2019}. The harder X-rays in 1.2--2.0~keV (black contours) concentrate at the base of the bubble on the galactic disk, which is probably associated with either the AGN or some nuclear starburst regions. There is a hard X-ray bright plume extending out of the nuclear region on the southwest side, which has been attributed to the synchrotron emission from the TeV CRs accelerated by the superbubble \citep{Li2019}. We herein only focus on the extended soft X-ray emission on the northeast (NE) side which is primarily from the hot gas.

\begin{figure*}%{r}{0.5\textwidth}
  \begin{center}
    \includegraphics[width=0.45\textwidth]{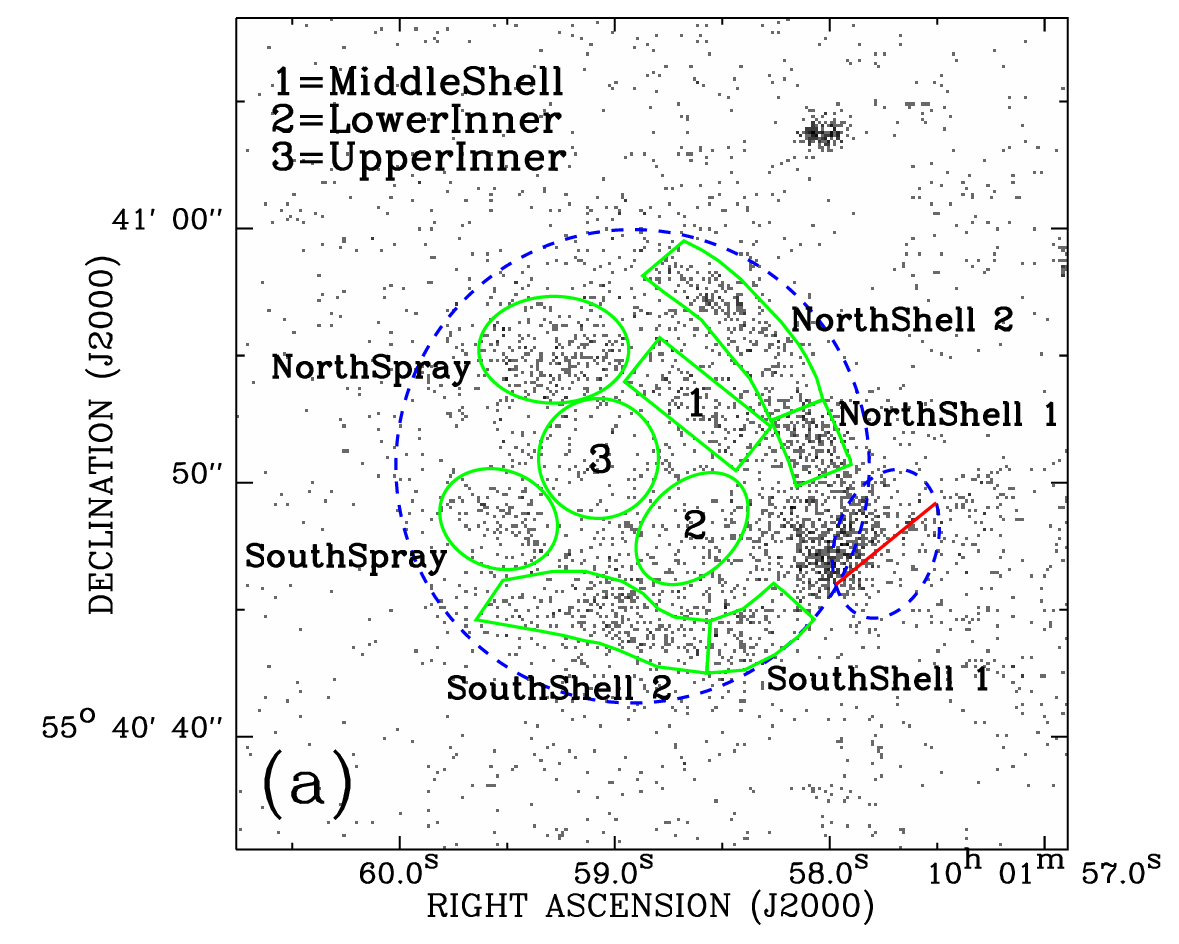}
    \includegraphics[width=0.37\textwidth,origin=br,angle=-90]{figures/nebubble_joint_apec_aspl.eps}
\caption{(a) Spectral extraction regions overlaid on the 0.5-2.0~keV \emph{Chandra} event map. The blue dashed circle is the entire NE superbubble, while the green solid regions are individual features of interest with their names marked beside. The blue dashed ellipse in the lower right is excluded as it may be affected by the scattered photons from the AGN. (b) Best-fit \emph{Chandra} spectra of the entire NE superbubble. The spectra extracted from the three observations are plotted in black, red, and green, respectively. }\label{fig:SpecRegions}
  \end{center}
\end{figure*}

We conduct spectral analysis of a few regions of prominent structures from the NE bubble (Fig.~\ref{fig:SpecRegions}a). Details of the spectral analysis, including the fitting models, uncertainties, and the derived physical parameters, are discussed in Appendix~\ref{sec:ChandraSpecAnal}. We herein only show an example of the spectra extracted from the entire NE superbubble after excluding the emission from the galactic nuclear region (Fig.~\ref{fig:SpecRegions}b). Apparently, the spectra from all the regions in the NE bubble is dominated by the thermal plasma with a characteristic temperature of $kT\sim1\rm~keV$ (Table~\ref{tab:XraySpecPara}; Fig.~\ref{fig:XraySpecIndividual}).

We also employ radio observations of NGC~3079 to probe the pressure of the magnetic field and the energy carried by the CR electrons in the superbubble. The radio data (Fig.~\ref{fig:MultiColorImage}a) is taken from the Continuum HAlos in Nearby Galaxies - an Evla Survey (CHANG-ES) project, observed with the Karl G. Jansky Very Large Array (VLA) at C-band (6\,GHz) in C- and D-configuration and at L-band (1.5\,GHz) in B-, C- and D-configuration. The details of this survey and data reduction is described in \citet{Irwin2012a,Irwin2012b,Wiegert2015}. The D- \citep{Wiegert2015} and B-configuration data \citep{Irwin2019} are public\,\footnote{CHANG-ES data releases available at www.queensu.ca/changes} and the C-configuration data release is in preparation (Walterbos et al. in prep).

We estimate the magnetic field strength of the NGC~3079 superbubble from the VLA radio observations. Estimating the magnetic field strength from the synchrotron emission requires an assumption on the energy equipartition between CRs and the magnetic field. This assumption is premised on an ideal scenario that CRs and magnetic field are strongly coupled and exchange energy until equilibrium is reached over a sufficient propagation time scale and length scale (typically $\sim1\rm~kpc$; e.g., \citealt{beckkrause2005,Seta2019}). This length scale is comparable to the size of the superbubble in NGC~3079, so there is a risk that the energy equipartition assumption may not be satisfied. Under the energy equipartition assumption, \citet{Li2019} estimated the magnetic field strength of $\approx23\rm~\mu G$ for the NE bubble, leading to a magnetic pressure of $\approx13\rm~eV~cm^{-3}$.

\subsection{SOFIA}\label{subsec:fir}

The \emph{SOFIA} observations of the nuclear region of NGC~3079 was taken in February and March of 2017 with the Field-Imaging Far-Infrared Line Spectrometer \citep[FIFI-LS; ][]{Klein2010,Fischer2018}. The FIFI-LS observations cover four strong emission lines from ionized gas in two different channels: [\ion{O}{3}]~$\lambda58\mu$m, [\ion{O}{3}]~$\lambda88\mu$m, [\ion{C}{2}]~$\lambda158\mu$m, [N{\small~II}]~$\lambda205\mu$m. The [\ion{O}{3}]~$\lambda58\mu$m and [\ion{N}{2}]~$\lambda205\mu$m lines have too low signal-to-noise ratio (S/N), so we only focus on the [\ion{O}{3}]~$\lambda88\mu$m and [\ion{C}{2}]~$\lambda158\mu$m lines in the present paper. The [\ion{O}{3}] line was covered in the blue channel with a field of view (FOV) of $\sim$30\arcsec$\times30$$^{\prime\prime}$ and a spectral resolution of $R\sim670$, while the [\ion{C}{2}] line was covered in the red channel (FOV $\sim1$\arcmin$\times$1\arcmin, $R\sim1200$). The integral field spectra of the [\ion{O}{3}] and [\ion{C}{2}] line combine over 30 and 50 exposures dithered in sub-pixel increments to boost the spatial sampling of the final data cube, which was resampled on a regular spatial grid with a pixel size of 1$^{\prime\prime}$ and 2$^{\prime\prime}$ for the [\ion{O}{3}] and [\ion{C}{2}] lines, respectively. The total effective exposure times are 1014~s and 1613~s for the two lines, respectively. The chop-subtraction, flat correction, telluric correction, flux/wavelength calibration, and spectral rebinning are performed following the instrument data reduction pipeline.

We use a Python GUI software SOSPEX \citep{Fadda2018} to  analyze the FIFI-LS data. We estimate the continuum level with a linear regression using the wavelength range with no significant line features on both sides of the emission lines. Such a continuum level is then subtracted from the data cube. We create the momentum maps by integrating the line emission in the velocity range of $\pm$600~km~s$^{-1}$ from the heliocentric velocity of NGC~3079 \citep[$v_{\rm helio}=1126$ km s$^{-1}$, ][]{Hagiwara2004}. The momentum maps are then cropped to discard the bad fittings with an intensity threshold of $10^{-18}$ erg s$^{-1}$ cm$^{-2}$ arcsec$^{-2}$ for [\ion{O}{3}] and $2.5\times10^{-18}$ erg s$^{-1}$ cm$^{-2}$ arcsec$^{-2}$ for [\ion{C}{2}]. The intensity maps of [\ion{O}{3}] and [\ion{C}{2}], as well as the velocity map of [\ion{C}{2}], are presented in Fig.~\ref{fig:MultiColorImage}b. Most of the far-IR line emissions concentrate in the nuclear region. As the highest-resolution [\ion{O}{3}]~$\lambda58\mu$m line is lost, we cannot resolve the bubble with the existing [\ion{O}{3}]~$\lambda88\mu$m and [\ion{C}{2}]~$\lambda158\mu$m lines. The shorter wavelength [\ion{O}{3}] line is only firmly detected in the nuclear region, while it is quite marginal to use the stronger [\ion{C}{2}] line to study the spatial distribution of the ionized gas within the superbubble (Fig.~\ref{fig:MultiColorImage}b; the angular resolution of SOFIA at 158~$\mu$m is FWHM $\sim$13\arcsec, comparable to the size of the superbubble; \citealt{Busch2018}). Nevertheless, we found a [\ion{C}{2}] velocity gradient of $\sim40$~km~s$^{-1}$~kpc$^{-1}$ on a larger scale along the major axis of the galaxy, which is consistent with a star forming rotating disk.

The [\ion{C}{2}] line emission is the dominant coolant of the neutral interstellar gas, so often adopted as one of the best extinction-free tracers of the obscured star formation. There are a few factors affecting the accuracy of taking the [\ion{C}{2}] line as the star formation tracer, such as the presence of an AGN, the IR color of the galaxy, and the efficiency of converting the far-UV radiation into gas heating, etc. (e.g., \citealt{HC2015}). For example, in principle, the ionic emission lines could also be largely produced by the AGN. We measure the fluxes of the [\ion{O}{3}] and [\ion{C}{2}] lines from a circular region with a radius of 13$^{\prime\prime}$ centered at the galactic nucleus. The relatively low line ratio of [\ion{O}{3}]/[\ion{C}{2}]$\sim0.1$ disfavors a strong contribution from the AGN \citep{HC2018a}. We then conclude that most of the line emission is produced by the starburst instead of the AGN. The [\ion{C}{2}] line flux in the central $r=13$$^{\prime\prime}$ circle is $1.5\times10^{41}\rm~ergs~s^{-1}$, corresponding to a star formation rate (SFR) of $1.3\rm~M_\odot~yr^{-1}$ \citep{HC2015}, which is consistent with the spatially resolved SFR estimate using the \emph{WISE} 22~$\mu$m photometry \citep{Vargas2018}.

\section{Discussion} \label{sec:dis}

\subsection{Pressure Balance between the Multiphase ISM in the Superbubble} \label{subsec:pres_comp}

In this section, we examine the balance between the thermal pressure of the hot and warm gases, the radiation pressure provided by young stars, and the CR and magnetic pressures in and around the bubble. As summarized in Table~\ref{tab:XraySpecPara}, the thermal pressure of the hot gas in different parts of the NE superbubbble is in the range of $(300-800)\rm~eV~cm^{-3}$, with a reasonable assumption on the volume filling factor $f_{\rm X}\sim1$. As the thermal pressure depends on $f_{\rm X}$ in the form of $P_{\rm X}\propto f_{\rm X}^{-1/2}$, the above thermal pressure is a lower limit under the resolution of current X-ray observations. On the other hand, the density of the warm ionized gas measured from the [\ion{S}{2}]~$\lambda\lambda6717,6731$ doublets in the bubble is $\lesssim10^2\rm~cm^{-3}$ \citep{Cecil2001}, which results in a thermal pressure $P_{\rm Opt}\lesssim100\rm~eV~cm^{-3}$, assuming a typical warm gas temperature of $T_{\rm e}\sim10^4\rm~K$. The thermal pressure of the hot gas could thus be about one order of magnitude higher than that of the warm ionized gas under current resolution of X-ray and optical observations.

We also made a rough estimate of the radiation pressure of the UV photons produced by the nuclear starburst. Adopting a SFR of $1.3~M_\odot$~yr$^{-1}$ and the stellar synthesis model (e.g., STARBURST99, \citealt{Leitherer1999}) with the metallicity of NGC~3079, we obtain an ionizing photon luminosity typically in the range of $L_{\rm ion}\sim(0.6-1.1)\times10^{43}$~erg~s$^{-1}$ for a continuous star-forming process. The uncertainties come from the adopted metallicity and stellar evolutionary models. If the ionizing photons are absorbed by the gas on the bubble shell and re-emitted isotropically in optical and infrared regime (without further process by the dust), the radiation pressure on a spherical bubble shell should be (e.g., \citealt{Lopez14}):
\begin{equation}
  P_{\rm rad}=\frac{L_{\rm ion,escape}}{4\pi{}r^2c}=\frac{f_{\rm esc}L_{\rm ion}}{4\pi{}r^2c}
\end{equation}
in which $f_{\rm esc}$ is the escaping fraction of the ionizing photons from the nuclear region. The radiation pressure from the nuclear starburst is then $\sim 10\rm~eV~cm^{-3}$ and $(0.4-0.8)\rm~eV~cm^{-3}$ at the base and top of the NE superbubble, assuming $f_{\rm esc}\sim1.0$. As $f_{\rm esc}$ is $<1$, the estimated radiation pressure is an upper limit, and significantly lower than the thermal pressure of the hot and warm gases. The above discussions do not consider the radiation pressure on the dust, which could be higher if the reprocessed IR emission is stronger. It is not quite clear how the dust is coupled with different gas phases and affects the dynamics of the superbubble.

As discussed in \S\ref{subsec:xray}, the magnetic field strength and the corresponding magnetic pressure of the NE bubble are $\sim23\rm~\mu G$ and $\sim13\rm~eV~cm^{-3}$, respectively. This is generally consistent with estimates from other literatures (e.g., \citealt{Li2019,Sebastian2019}). Apparently, the magnetic pressure is lower than the thermal pressure of both hot and warm ionized gases. However, a direct comparison between them may be more complicated, as there is no one-to-one correlation between specific radio/X-ray/optical emitting structures. The radio lobe is significantly more extended than the X-ray and optical emission line bubbles (Fig.~\ref{fig:MultiColorImage}), so we cannot rule out the possibility that there is a pressure balance between them at the radio ``cap''. Considering the overall spatial distribution of the multi-wavelength emissions from the NE side (radio ``cap'' above the X-ray and optical emission line ``bubble''), we conclude that the magnetic pressure could at least play a role (if not dominant) in the overall dynamics of the galactic nuclear superbubble in NGC~3079.

The above examination of the pressure balance in the NGC~3079 superbubble could be compared to those in active star forming regions, such as 30~Doradus in the Large Magellanic Cloud. \citet{Pellegrini2011} found that in most part of 30~Doradus the $P_{\rm hot}/P_{\rm warm}$ ratio is $\sim1-10$, while the radiation pressure is generally not important compared to the thermal pressures (also see \citealt{Barnes2020}). Combined with our analysis for NGC~3079, we then conclude that in star formation regions of scale less than kpc (including the galactic nuclear region), the gas expansion and outflow is mainly thermal pressure driven, with the possibility of being affected by the magnetic pressure.

\subsection{Energy Budget of the Superbubble}\label{subsec:dyn}

In this section, we further compare the energy detected in the superbubble to different energy sources from the galaxy. According to our X-ray spectral analysis, the total thermal energy of the hot gas in the NE superbubble is $E_{\rm th,hot}\sim 1.4\times10^{55}~f_{\rm X}^{1/2}\rm~ergs$. The hot gas radiative cooling timescale is typically in the range of $\tau_{\rm cool}\approx(2-7)\times10^7\rm~yr$ of the resolved features (Table~\ref{tab:XraySpecPara}; even longer for unresolved lower surface brightness features), which is longer than the dynamical age of $\lesssim1\rm~Myr$ of the superbubble (see below). Therefore, the radiation loss of the hot gas is negligible during the evolution of the superbubble. For comparison, \citet{Cecil2001} reported a kinetic energy of the warm ionized gas of $E_{\rm k,warm}\sim3\times10^{54}~f_{\rm warm}^{1/2}\rm~ergs$, where $f_{\rm warm}$ is the volume filling factor of the warm ionized gas. We then expect $E_{\rm k,warm}<<E_{\rm th,hot}$, especially considering that the optical emission line features may have $f_{\rm warm}<<1$. However, cautions should be made that we may have over estimated $E_{\rm th,hot}$, as a significant fraction of the soft X-ray emission may come from non-thermal processes such as the charge exchange \citep{Zhang2014}, although these processes may not qualitatively change our conclusions on the energy budget.

Consistent with the above estimates, we then compare our measurements to the analytical model of the superbubble evolution in the intermediate stage, developed by \citet{Castor75} and \citet{Weaver77}. This model accounts for a steady mechanical energy injection, and the growth of the bubble is driven by the adiabatic expansion of the shock heated hot gas. The shocked ambient medium forms a dense shell around the hot bubble under the circumstance of rapid cooling. The age of the bubble in the adiabatic expansion stage is linked to the current radius and velocity in the following form:
\begin{eqnarray}
    t_{\rm Myr} = 0.59 R_{\rm kpc}/v_{\rm exp,3},
\end{eqnarray}
where $t_{\rm Myr}$, $R_{\rm kpc}$, and $v_{\rm exp,3}$ are the age in Myr, radius in kpc, and expansion velocity in $10^3\rm~km~s^{-1}$, respectively. Adopting the velocity measured via the shift of optical emission lines [typcially $\sim(500-1000)\rm~km~s^{-1}$); \citealt{Veilleux1994}] and a radius of $R=0.75\rm~kpc$, we obtain the range of the age of the bubble $t\sim(0.4-0.8)\rm~Myr$. On the other hand, the hot bubble model predicts that the thermal energy of the hot gas takes 5/11 of the total mechanical energy \citep{MM88}. Therefore, the observed total amount of thermal energy of hot gas requires an average energy injection rate of $\dot{E}_{\rm hot}\approx(1.2-2.5)\times10^{42}\rm~ergs~s^{-1}$.

We then estimate the energy provided by nuclear starburst and AGN. We use the stellar synthesis model STARBURST99 to estimate the mechanical energy injection rate in the case of continuous star-forming process. At a SFR of $\sim1.3~M_\odot$~yr$^{-1}$ for the nuclear region of NGC~3079 (\S\ref{subsec:fir}), the stellar synthesis model predicts a growing mechanical energy injection rate of $\dot{E}_{\rm SF}$ due to the accumulation of SN events at the early stages. $\dot{E}_{\rm SF}$ stabilizes at $\sim(5-6)\times10^{41}\rm~ergs~s^{-1}$ at the age of $\gtrsim~40\rm~Myr$. This $\dot{E}_{\rm SF}$ could thus be adopted as an upper limit of the SF energy injection, which is still below the required energy injection we estimated above, let alone that the cold gas may carry kinetic energy of similar amount \citep[e.g.,][]{Cecil2001}. Besides that, the energy loss due to radiation, thermal conduction, or even break-out at the top of the superbubble may consume a significant fraction of the injected mechanical energy. Therefore, we conclude that current nuclear starburst is not powerful enough to drive the expansion of the superbubble.

The parsec-scale jet revealed by radio VLBI observations may provide additional power to drive the superbubble. \citet{Shafi2015} reported a power of $\dot{E}_{\rm jet}\sim4\times10^{41}$~ergs~s$^{-1}$ for the parsec-scale jet, according to its radio continuum luminosity and empirical relations \citep[e.g.,][]{2008ApJ...686..859B,2010ApJ...720.1066C}. This jet power, however, is still not enough to match the requirement of the observed hot gas thermal energy of the superbubble. If neither the current nuclear starburst or the pc-scale jet can be energetic enough to account for the observed hot gas thermal energy, we speculate that the past AGN activity may help to inject more mechanical energy to drive the expansion of the nuclear superbubble. Such a scenario has also been employed to explain the ``Fermi bubble'' in the MW (e.g., \citealt{Guo2012}), and supported by the discovery of the ionization cones toward the Magellanic stream which could be a fossil imprint of a powerful flare from the MW center \citep{BH2013}.

\section{Conclusion} \label{sec:con}

Based on the multi-wavelength data of the bipolar kpc-scale nuclear superbubble in NGC~3079, we examined the pressure balance between different ISM phases. We found that the thermal pressure of the X-ray emitting hot gas is about one order of magnitude higher than that of the optical line emitting warm ionized gas, and both of them are far larger than the radiation pressure provided by the nuclear starburst. The magnetic pressure is apparently lower, but has a significantly different spatial distribution, so may still be important in the bubble expansion, especially on larger scales.

The observed thermal energy of the hot gas enclosed by the bubble requires an average energy injection rate within the dynamical age of the bubble significantly larger than those expected from the nuclear starburst and the pc-scale jet. Additional energy sources, such as past activities of the AGN, may be possible solutions to be at least partially responsible for driving the bubble. How the AGN and nuclear starburst convert their energy to the thermal energy (and other forms of energy), and how they drive the expansion of the bubble, remain to be investigated with future observational and theoretical works.

\acknowledgments

The authors acknowledge Dr. Yelena Stein from the German space agency for helpful discussions on the calculation of the magnetic field strength. J.T.L. and Y.Y. acknowledge the financial support from the National Science Foundation of China (NSFC) through the grant 12273111. Y.Y. acknowledges support from the National Natural Science Foundation of China (NSFC) through the grant 12203098. This paper employs a list of Chandra datasets, obtained by the Chandra X-ray Observatory, contained in~\dataset[DOI: 10.25574/cdc.226]{https://doi.org/10.25574/cdc.226}.

\appendix
\section{Chandra Spectral Analysis}\label{sec:ChandraSpecAnal}

In this section, we present details of the \emph{Chandra} spectral analysis of individual regions inside the NE superbubble of NGC~3079 (Fig.~\ref{fig:SpecRegions}a). All the source spectra are grouped to a $\rm S/N\gtrsim3$ in each bin. The sky background spectra are extracted from a region without significant emission on the same CCD chip. The spectra of each region extracted from the three observations are jointly analyzed with XSPEC.

All the spectra are fitted with a single absorbed (TBabs) thermal plasma model (APEC). The absorption contains two components: the Milky Way foreground one is fixed at an average value in the direction toward NGC~3079, i.e., $N_{\rm H}=8.74\times10^{19}$~cm$^{-2}$. Furthermore, some regions may be highly impacted by the intrinsic absorption from the cold gas in the disk of NGC~3079. This component is allowed to vary. We notice that except for the regions close to the galaxy disk (``NorthShell 1'', ``SouthShell 1'', and ``LowerInner''), this additional absorption component is often negligible within the uncertainty. This indicates that the NE side is the foreground side to the observer, while the opposite side, which suffer from stronger absorption \citep{Li2019}, is in the background.

When fitting the spectra from the entire NE bubble (Fig.~\ref{fig:SpecRegions}b), the relative abundance to the solar value \citep{Asplund2009} is also set free. However, the abundance is often poorly constrained in smaller regions with poorer counting statistic, so we fix it at the best-fit value of the entire bubble ($Z\approx 0.16\rm~Z_\odot$). The best-fit spectra from individual regions is presented in Fig.~\ref{fig:XraySpecIndividual}, while the corresponding best-fit model parameters are summarized in Table~\ref{tab:XraySpecPara}.

We also derive some physical parameters of the hot gas from the directly fitted model parameters. The electron number density $n_e$ is derived from the best-fit emission measure (EM), assuming a $n_e$ to $n_{\rm H}$ ratio of 1.2. When calculating the density, we need to estimate the volume of the spectral analysis regions. We assume an axisymmetric geometry of all these regions. We adopt an oblate ellipsoid shape of the ``LowerInner'', ``UpperInner'', ``NorthSpray'', and ``SouthSpray'' regions, and a cylinder shape with a height of 560~pc and a radius of 110~pc of the ``MiddleShell'' region. For the four ``Shell'' regions, ``NorthShell 1'', ``NorthShell 2'', ``SouthShell 1'', and ``SouthShell 2'', we assume they are part of a spherical region with a diameter of 1.5~kpc and a thickness of 220~pc (2.2\arcsec). The actual volume occupied by the hot gas is also characterized with the volume filling factor $f_{\rm X}$, which is defined as the fraction of an unresolved volume occupied by the hot gas. Since the hot gas typically has the highest specific energy or pressure among different ISM phases, the typical value of the volume filling factor is often $f_{\rm X}\sim1$, so does not affect our scientific discussions. The average thermal pressure ($P_{\rm th}$) and total thermal energy ($E_{\rm th}$) of hot gas within the regions are then derived from $n_e$ and the best-fit plasma temperature $kT$. The cooling timescale of the X-ray emitting gas is: $\tau_{\rm cool}\approx1.3\times10^5T^{1.7}_6/n_e$, where $T_6$ is the plasma temperature in $10^6$~K and $n_e$ in $\rm cm^{-3}$ \citep{Draine2011}. We present the confidence contours of EM, $kT$, and the thermal pressure of individual regions also in Fig.~\ref{fig:XraySpecIndividual}. The derived parameters are also listed in Table~\ref{tab:XraySpecPara}. 

\begin{table*}
%\vspace{-0.in}
\begin{center}
\rotatebox{90}{
\begin{minipage}{\textheight}
\caption{Parameters of the hot gas component from individual regions in the NE bubble of NGC~3079}\label{tab:XraySpecPara}
\setlength{\leftskip}{-40pt}
%\tiny
%\scriptsize
\footnotesize
%\hspace{-1in}
\begin{tabular}{lcccccccccccl}
\hline\hline
 \colhead{} & \colhead{$N_{\rm H,NGC3079}$} & \colhead{$kT$} & \colhead{Abundance} & \colhead{EM} & \colhead{$\chi^2/d.o.f.$} & \colhead{$V$} & \colhead{$n_e$} & \colhead{$P_{\rm th}$} & \colhead{$E_{\rm th}$} & \colhead{$\tau_{\rm cool}$} \\
 \colhead{Spectral Region} & \colhead{($10^{21}$cm$^{-2}$)} & \colhead{(keV)} & \colhead{($Z_\odot$)} & \colhead{($10^{62}\rm cm^{-3}$)} & \colhead{} & \colhead{($10^{63}\rm cm^3$)} & \colhead{($f^{-1/2}_{\rm X}\rm cm^{-3}$)} & \colhead{($10^{2}f^{-1/2}_{\rm X}\rm eV cm^{-3}$)} & \colhead{($10^{54}f^{1/2}_{\rm X}\rm ergs$)} & \colhead{(10$^7f^{1/2}$ yr)}\\
 \hline
 NE superbubble\dotfill & $0.4\pm0.3$ & $0.88_{-0.04}^{+0.03}$ & $0.16_{-0.03}^{+0.04}$ & $6.0_{-1.2}^{+1.5}$ & 131.62/132 & \nodata & \nodata & \nodata & \nodata & \nodata \\
 NorthShell 1\dotfill & $1.6_{-1.4}^{+2.2}$ & $0.98_{-0.20}^{+0.17}$ & 0.16(fixed) & $0.88_{-0.22}^{+0.51}$ & 18.71/12     & 1.4  & $0.28_{-0.04}^{+0.07}$ &
 % $8.4\pm1.3$ 
 $5.2\pm0.8$
 & $1.8\pm0.3$ & $2.9\pm1.8$ \\
 SouthShell 1\dotfill & 0.7($<6.8$)  & $1.10_{-0.59}^{+0.32}$ & 0.16(fixed)  & $0.44_{-0.11}^{+1.63}$ & 2.47/7       & 1.9  & $0.17_{-0.02}^{+0.19}$ &
 % $5.6_{-1.4}^{+1.8}$
 $3.5_{-0.8}^{+1.1}$
 & $1.6_{-0.4}^{+0.5}$ & $5.7_{-4.9}^{+5.9}$ \\
 LowerInner\dotfill       & 3.1($<8.3$)  & $0.66_{-0.36}^{+0.44}$ & 0.16(fixed)  & $0.87_{-0.53}^{+7.31}$ & 2.47/7       & 1.4  & $0.27_{-0.10}^{+0.56}$ &
 % $5.5_{-1.2}^{+3.3}$
 $3.4_{-0.8}^{+2.0}$
 & $1.2_{-0.3}^{+0.7}$ & $1.5_{-1.3}^{+2.5}$ \\ 
 NorthShell 2\dotfill     & $<1.0$       & $0.80_{-0.11}^{+0.09}$ & 0.16(fixed)  & $0.96_{-0.11}^{+0.32}$ & 19.27/18     & 4.2  & $0.17_{-0.01}^{+0.03}$ &
 % $4.1_{-0.5}^{+0.6}$
 $2.6_{-0.3}^{+0.4}$
 & $2.6\pm0.3$         & $3.4_{-1.3}^{+1.1}$ \\
 MiddleShell\dotfill      & $<0.80$      & $0.86\pm0.10$          & 0.16(fixed)  & $0.55_{-0.07}^{+0.12}$ & 11.00/10     & 0.6  & $0.32_{-0.02}^{+0.03}$ &
 % $8.6_{-1.4}^{+1.3}$
 $5.4_{-0.9}^{+0.8}$
 & $0.8\pm0.1$         & $2.0\pm0.7$ \\
 UpperInner\dotfill       & $<7.3$       & $0.71_{-0.48}^{+0.29}$ & 0.16(fixed)  & $0.27_{-0.07}^{+9.14}$ & 4.73/4       & 1.6  & $0.14_{-0.02}^{+0.04}$ &
 % $3.1_{-0.9}^{+1.5}$
 $1.9_{-0.5}^{+0.9}$
 & $0.7_{-0.2}^{+0.4}$ & $3.2_{-2.3}^{+4.5}$ \\
 SouthShell 2\dotfill     & $<0.81$      & $0.81_{-0.09}^{+0.08}$ & 0.16(fixed)  & $0.99_{-0.11}^{+0.25}$ & 16.33/17     & 4.5  & $0.16_{-0.01}^{+0.02}$ &
 % $4.0\pm0.5$
 $2.5\pm0.3$
 & $2.7\pm0.3$         & $3.6_{-1.2}^{+1.0}$ \\
 NorthSpray\dotfill       & $<0.46$      & $0.81_{-0.09}^{+0.10}$ & 0.16(fixed)  & $0.93\pm0.11$          & 20.96/17     & 2.2  & $0.22\pm0.01$          &
 % $5.6\pm0.8$ 
 $3.5\pm0.5$
 & $1.9\pm0.3$         & $2.6_{-0.7}^{+0.8}$ \\
 SouthSpray\dotfill       & $<2.3$       & $0.80_{-0.21}^{+0.14}$ & 0.16(fixed)  & $0.46_{-0.07}^{+0.44}$ & 13.29/9      & 1.3  & $0.20_{-0.02}^{+0.08}$ &
 % $5.0_{-1.0}^{+1.1}$
 $3.1_{-0.6}^{+0.7}$
 & $1.0\pm0.2$         & $2.8\pm1.5$ \\
\hline\hline\\
\end{tabular}
\normalsize
Parameters of hot gas obtained from X-ray spectral analysis of individual regions (Figs.~\ref{fig:SpecRegions}, \ref{fig:XraySpecIndividual}). $N_{\rm H,NGC3079}$ is the additional absorbing Hydrogen column density from the local ISM of NGC~3079, above the MW value of $N_{\rm H}=8.74\times10^{19}$~cm$^{-2}$. EM is the directly measured hot gas emission measure, defined as: EM=$n_en_{\rm H}f_{\rm X}V$, where $f_{\rm X}$ is the volume filling factor within the volume $V$. $n_e$, $P_{\rm th}$, $E_{\rm th}$, and $\tau_{\rm cool}$ are the derived electron number density, hot gas average thermal pressure, total thermal energy and hot gas cooling timescale.
\end{minipage}
}
\end{center}
\end{table*}

\begin{figure*}
  \includegraphics[width=0.37\textwidth,origin=br,angle=-90]{figures/southshell2_jf_z0.16_apec.eps}
  \includegraphics[width=0.50\textwidth]{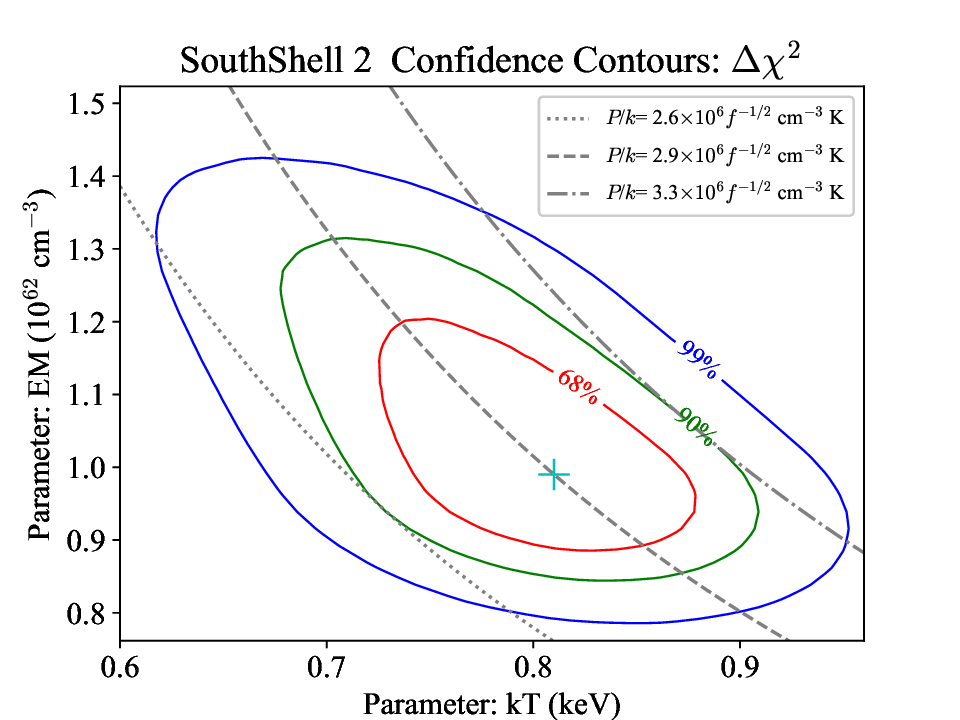} \, 
  \includegraphics[width=0.37\textwidth,origin=br,angle=-90]{figures/northspray_jf_z0.16_apec.eps}
  \includegraphics[width=0.50\textwidth]{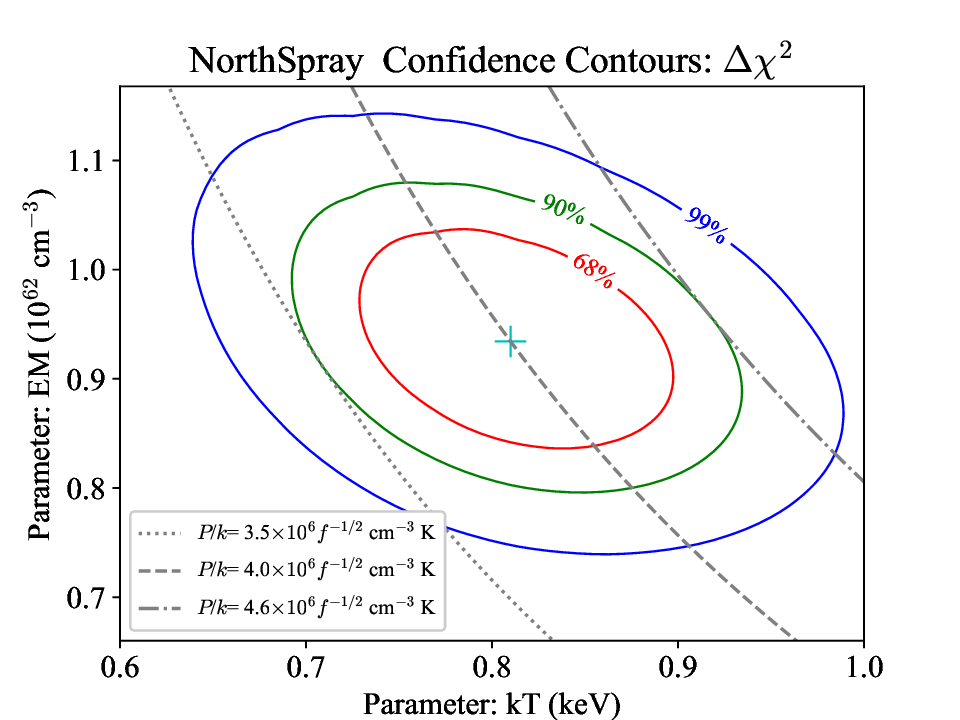} \,
\includegraphics[width=0.37\textwidth,origin=br,angle=-90]{figures/southspray_jf_z0.16_apec.eps}
  \includegraphics[width=0.50\textwidth]{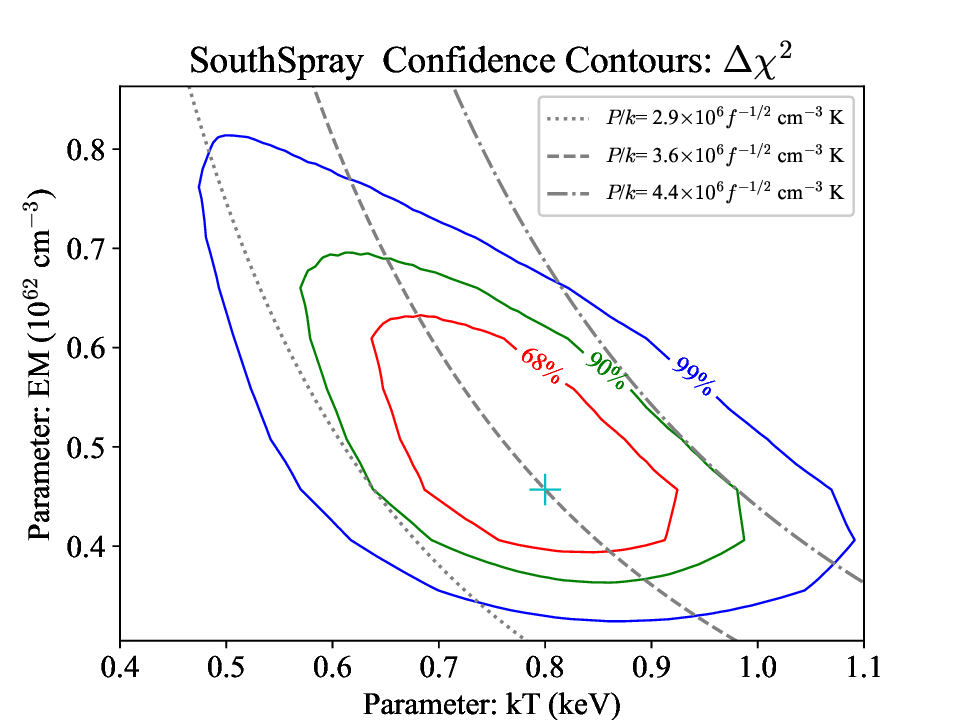} \,
\caption{\emph{Chandra} spectra extracted from individual regions in the NE bubble of NGC~3079 (left) and the confidence contours of the plasma temperature ($kT$) and emission measure (EM) of the best-fit models (right). Different colors in the left panels denote data and model of the spectra extracted from the three different observations, which are jointly fitted together. In the right panel, the best-fit result is marked as the cyan cross, and the red, green, and blue contours are at levels of $\Delta\chi^2=2.30$, 4.61, and 9.21, corresponding to the confidence area of 68\%, 90\%, and 99\% possibilities. The lower and upper limits of the hot gas thermal pressure are determined using the isobaric curves tangent to the 90\% confidence contour (dotted and dot-dashed lines).}\label{fig:XraySpecIndividual}
\end{figure*}
\addtocounter{figure}{-1}
\begin{figure}
\centering
\includegraphics[width=0.37\textwidth,origin=br,angle=-90]{figures/northshell1_jf_z0.16_apec.eps}
  \includegraphics[width=0.50\textwidth]{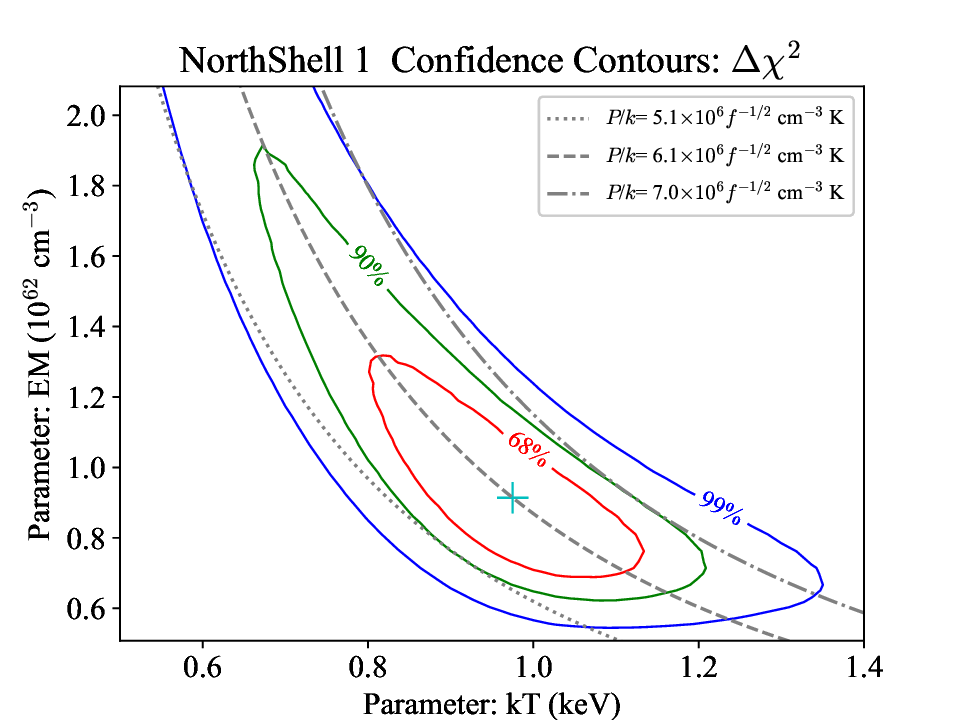} \,
\includegraphics[width=0.37\textwidth,origin=br,angle=-90]{figures/southshell1_jf_z0.16_apec.eps}
  \includegraphics[width=0.50\textwidth]{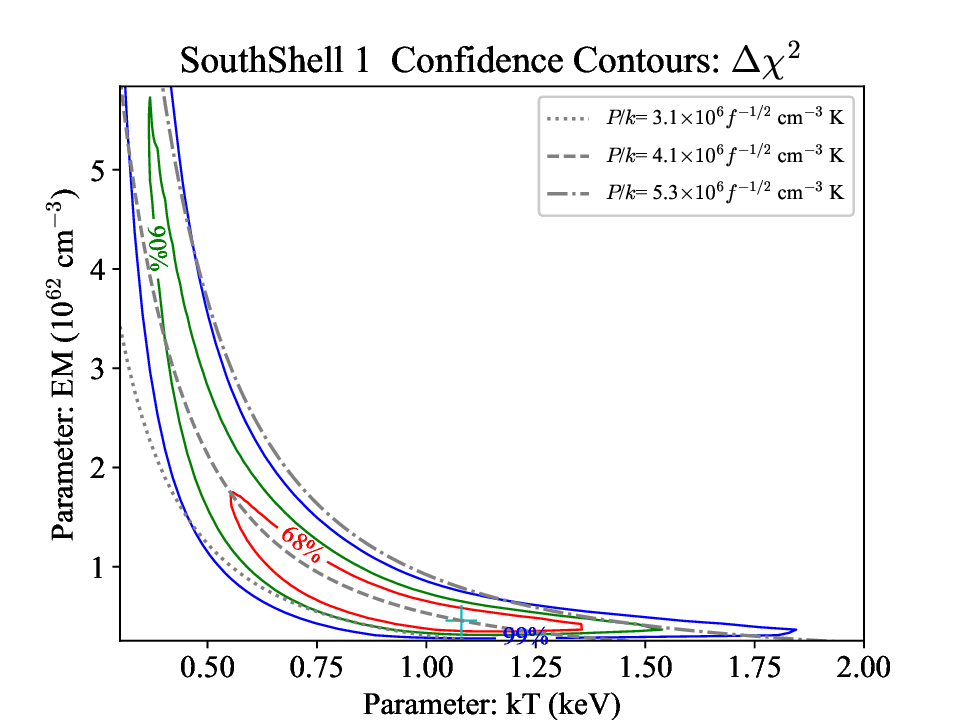} \,
\includegraphics[width=0.37\textwidth,origin=br,angle=-90]{figures/lowerinner_jf_z0.16_apec.eps}
  \includegraphics[width=0.50\textwidth]{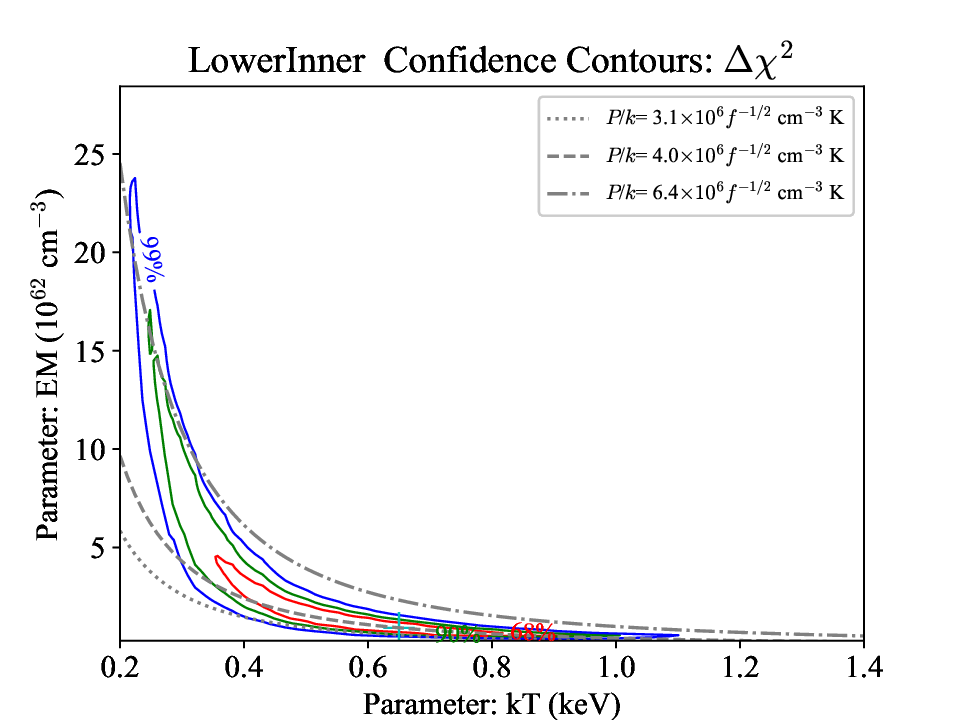}  \,
\caption{continued}
\end{figure}
\addtocounter{figure}{-1}
\begin{figure}
\centering
\includegraphics[width=0.37\textwidth,origin=br,angle=-90]{figures/northshell2_jf_z0.16_apec.eps}
  \includegraphics[width=0.50\textwidth]{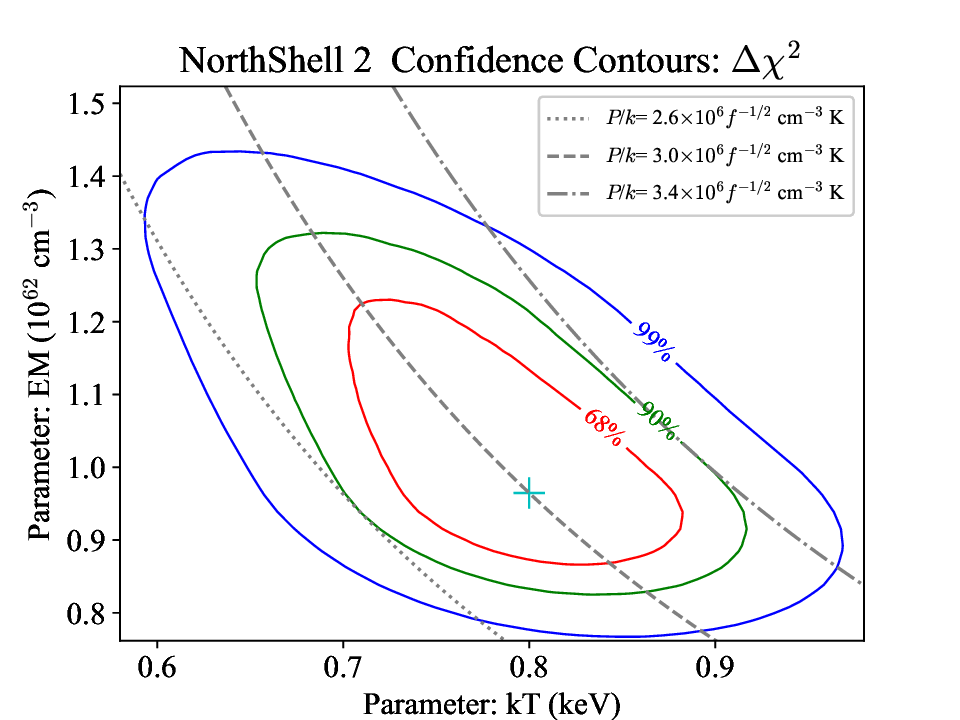} \,
\includegraphics[width=0.37\textwidth,origin=br,angle=-90]{figures/middleshell_jf_z0.16_apec.eps}
  \includegraphics[width=0.50\textwidth]{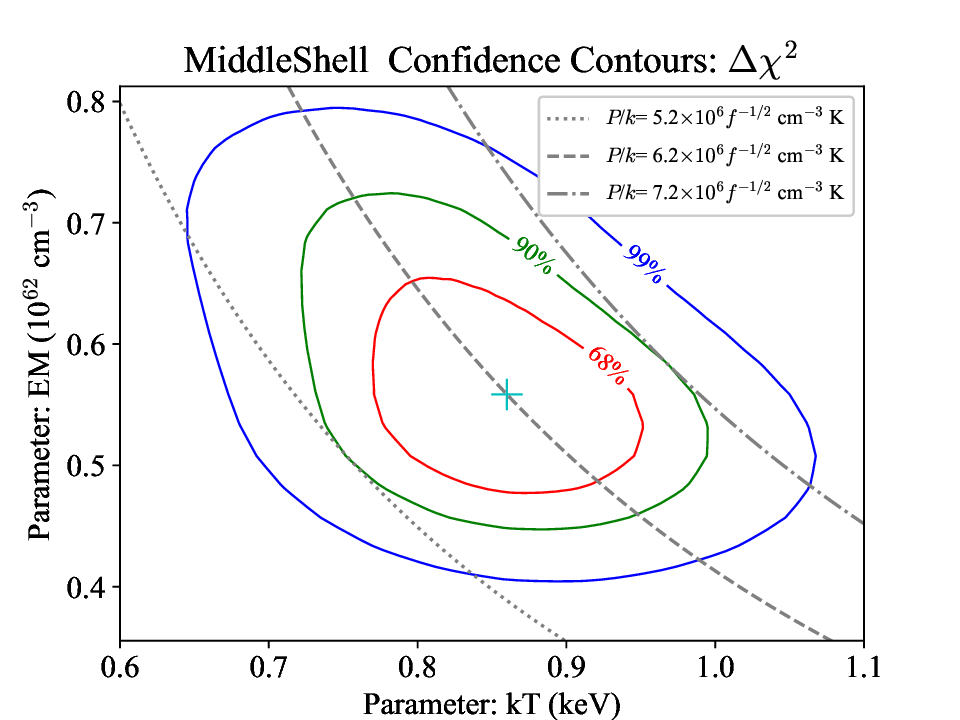} \,
\includegraphics[width=0.37\textwidth,origin=br,angle=-90]{figures/upperinner_jf_z0.16_apec.eps}
  \includegraphics[width=0.50\textwidth]{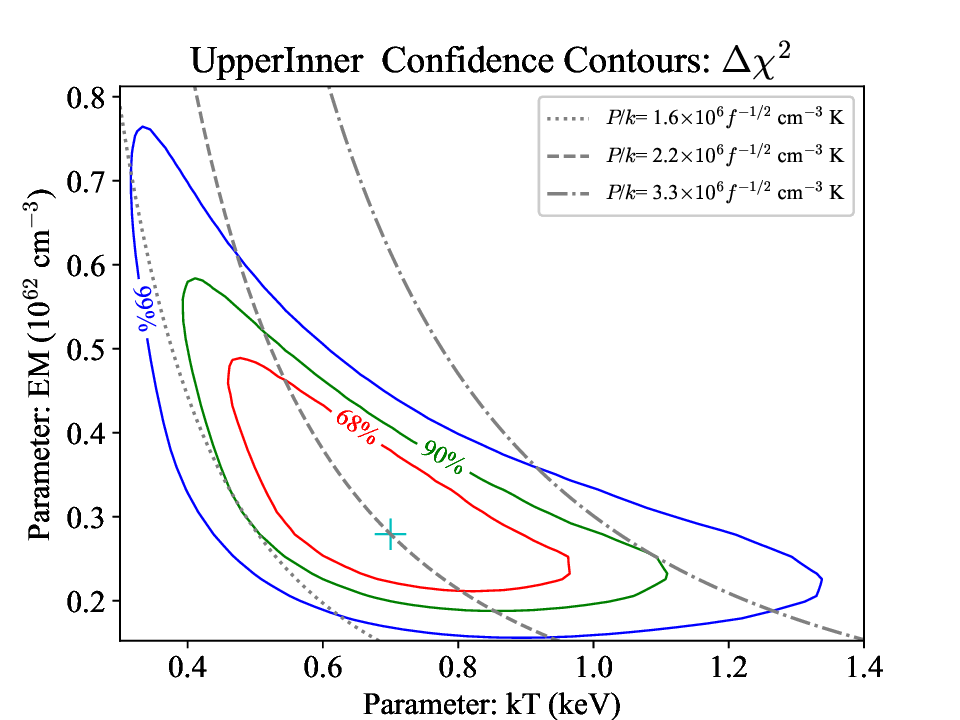}  \,
\caption{continued}
\end{figure}

\bibliography{references}
\bibliographystyle{aasjournal}

\end{document}